\documentstyle[12pt]{article}
\begin{document}
\thispagestyle{empty}   
\begin{flushleft}

Oslo TP 17-93\\

ZTF - 93/11\\

\end{flushleft}
\vspace{1.0cm}

\begin{center}

  \begin{Large}

  \begin{bf}

New contribution to anomalous radiative $K^0$-decays

  \end{bf}

  \end{Large}

\vspace{1cm}

  \begin{Large}

J. O. Eeg\footnote{e-mail : eeg@vuoep1.uio.no}\\
\vspace{0.5cm}

{\small Dept. of Physics, University of Oslo,
N-0316 Oslo, Norway\\}
\vspace{1.0cm}

 I. Picek\footnote{e-mail : picek@phy.hr}\\
\vspace{0.5cm}

{\small Fizi\v{c}ki odjel, University of Zagreb,
POB 162, 41000 Zagreb, Croatia\\}

\end{Large}

\end{center}

\vspace{0.7cm}
\begin{center}
  {\bf Abstract}
\end{center}
\begin{quotation}
\noindent
We present a new contribution to the
 $K_L\rightarrow \gamma \gamma$ amplitude,
which is ${\cal{O}}(p^4)$ within
the counting rules of chiral perturbation theory.
This direct (non-pole) amplitude, obtained from short-distance
$s \rightarrow d \gamma$
 quark diagrams of order $e G_F \alpha_s/\pi$,
and similar $s \rightarrow d \gamma \gamma$ diagrams,
can  account for about
 half of the experimental amplitude. Closely following  the description of the
 $\pi ^0\rightarrow \gamma \gamma$ and $K_L\rightarrow \gamma \gamma$
processes in the variants of the same low-energy QCD, we find that both of
these processes are anomalous  in the same sense. Then, by the anomaly-matching
principle, we arrive from the chiral-quark to the bosonic counterparts for both
of these processes.
In this way we add the $K_L\rightarrow \gamma \gamma$ decay to
 the existing list of anomalous radiative prosesses.

\end{quotation}

\newpage

\begin{Large}

1. Introduction

\end{Large}

\vspace{0.2cm}

Radiative kaon decays appear to be  suitable for uncovering the
subtleties of overbridging the quark and the hadronic worlds. This
procedure conventionally derives the effective bosonic operators of the
Chiral Perturbation Theory ($\chi$PT),
 starting from {\em four-quark} operators at the free-quark level
\cite{PdeR91}.

In this paper we recall our recent experience in transferring a
{\em two-quark} operator  at quark level to a contribution to the
$\overline{K^0} \rightarrow \gamma \gamma$ amplitude \cite{ep93}.
In an earlier paper \cite{sdgg} we pointed
out the existence of pure electroweak short-distance
$s \rightarrow d \gamma \gamma$ loop diagrams
that gave an important contribution to the CP-violating
$\overline{K^0} \rightarrow \gamma \gamma$ amplitude.
The significance of the two-quark
$s \rightarrow d \gamma \gamma$ operator was due to  the non-efficient
GIM cancellation for a heavy top-quark in the relevant loop diagrams.
Completing these irreducible diagrams \cite{sdgg} with the reducible ones
gave large cancellations for free on-shell $s,d$ quarks \cite{red}.
Using an effective low-energy QCD advocated by many authors \cite{leqcd}, we
transformed the short distance
 quark result into the meson-"world" amplitude \cite{ep93},
for which these cancellations were lifted.
Our  new $\overline{K^0} \rightarrow \gamma \gamma$
amplitude obtained in this way is   ${\cal{O}}(p^4)$ within the counting rules
of  $\chi$PT. However,  to our knowledge, there  are
no such direct  (non-pole)  $\chi PT$ terms
${\cal{O}}(p^4)$  in the literature.

In the present paper we dwell on the appearance of this
$\overline{K^0} \rightarrow \gamma \gamma$ amplitude at the
hadronic level. Thereby we rely  on the close relation which we demonstrate
to exist between the
 $\pi ^0\rightarrow \gamma \gamma$ and $K_L\rightarrow \gamma \gamma$
amplitudes. Whereas the former is known to be governed by
the anomalous Wess-Zumino-Witten (WZW) \cite{WZ} term, which
is ${\cal{O}}(p^4)$ , the existing  locally chiral-invariant
flavour-changing
contributions  of ${\cal{O}}(p^4)$ considered in  \cite{enp,bep,p92}
correspond to four-quark operators, and do not contribute to
$\overline{K^0} \rightarrow \gamma \gamma$.  The anomalous
terms presented by these authors \cite{enp,bep,p92},
termed the {\em direct anomalous} ones, can be constructed from a
combination of the anomalous current obtained from the WZW term
 (responsible for $\pi^0 \rightarrow \gamma \gamma$) and the  current
obtained from  the normal term ${\cal{O}}(p^2)$.
The {\em reducible anomalous} pole diagrams, including directly the WZW term
for $\pi^0, \eta  \rightarrow 2 \gamma$,
vanish in the SU(3) limit where the Gell-Mann-Okubo
mass formula is used. The remaining reducible anomalous contributions
to $\overline{K^0} \rightarrow \gamma \gamma$ come from
 ${\cal{O}}(p^6)$ corrections to the Gell-Mann-Okubo
mass formula.

The purpose of this paper is twofold: First, we present
a CP-conserving counterpart of ref. \cite{sdgg}. Such a
 CP-conserving $\overline{K^0} \rightarrow \gamma \gamma$
amplitude, suffering from the GIM cancellation
in the pure electroweak case, has to be of
order $e^2 G_F \alpha_s/\pi$, as shown in some of the early
literature on this process\cite{old}. Second, we
demonstrate  the anomalous nature of this process (in the low-energy
QCD at hand) and
discuss how   our  ${\cal{O}}(p^4)$,
$\overline{K^0} \rightarrow \gamma \gamma$ amplitude,
 corresponding to two-quark operators,
 can be included among $\Delta S = 1$ chiral lagrangian terms.

\vspace{0.5cm}

\begin{Large}

2. The quark evaluation of the $K_L \rightarrow  \gamma \gamma$ amplitude

\end{Large}

\vspace{0.2cm}

 In previous papers we studied electroweak quark loops for
$s \rightarrow d \gamma$ and
$s \rightarrow d \gamma \gamma$ \cite{sdgg,red},
 where the parts related by Ward identities can be combined into
 the effective lagrangian
\begin{equation}
{\cal{L}}(s\rightarrow d)_{\gamma} \, = \,  B \,
\epsilon^{\mu \nu \lambda \rho}
F_{\mu \nu} \, ( \overline{d_L} \; i \stackrel{\leftrightarrow}{D_{\lambda}}
\; \gamma_{\rho} s_L )  \; .
\end{equation}
The quantity $B$ contains short distance loop effects from scales
above the chiral symmetry breaking scale $\sim$ 1 GeV. $F_{\mu \nu}$
is the electromagnetic field tensor.
The covariant derivative $\stackrel{\leftrightarrow}{D_{\lambda}}$,
containing another electromagnetic field $A_\lambda$,
acts on left-handed  $s$- and $d$- quark fields.
Having obtained ${\cal{L}}(s\rightarrow~d)_{\gamma}$,  one often
removes the covariant derivative by using the equations of
motion. Then one
obtains the well-known magnetic moment term\cite{SVZg}.
In other words,  one can split ${\cal{L}}(s\rightarrow d)_{\gamma}$
in two terms (eqs. (9) and (10) of ref.\cite{ep93}),
the one which vanishes at
the free-quark  mass-shell and the other (the magnetic moment term
of  ${\cal{O}}(p^6)$ in $\chi PT$) that vanishes in the  limit
$m_{s,d} \rightarrow 0$.

The quantity $B$ is
of order $e \cdot G_F$ and, for convenience, let us introduce a
dimensionless quantity $\hat{B}$ :
\begin{equation}
 B \, = \, \frac{G_F}{\sqrt{2}} \frac{e}{4 \pi^2}\;  \hat{B}\; .
\end{equation}
Furthermore,  $\hat{B}$ can be split in two parts
with different KM-factors
 $\lambda_q = V_{qs} V_{qd}^* \, (q=u,c,t)$ ,
\begin{equation}
\hat{B} \, = \, \lambda_u \, \hat{B}_u \, +
\,  \lambda_t \, \hat{B}_t \,  .
\end{equation}
The first term is purely CP-conserving (CP-even), whereas the
second contributes to the CP-violating effects determined by
$Im(\lambda_t)$.

In the present theoretical approach, the
 $\overline{K^0} \rightarrow  \gamma \gamma$
amplitude is calculated  within a low-energy QCD model from  quark-level
diagrams.
 This low-energy QCD \cite{leqcd}
 is obtained by adding a new term to ordinary QCD:
\begin{equation}
  {\cal{L}}_{\chi} = - M (\overline{q_R} \; U q_L +
\overline{q_L} \; U^\dagger q_R) \; ,
\end{equation}
where $\bar{q} = (\bar{u},\bar{d},\bar{s})$ and the 3 by 3 matrix
$U \equiv \exp\biggl({{2i\over f}\Pi}\biggr)$
contains the pseudoscalar octet mesons
 $\Pi = \sum_a \pi^a\lambda^a/2 \,$
$(a=1,..,8)$, and
$f$ can be identified with the pion decay constant,
$f = f_\pi = (92.4 \pm 0.2)~{\rm MeV}$ ($=f_K,$ in the chiral limit).
The resulting field theory with
pseudoscalar quark-meson couplings and quark loops reproduces the
well-known $\pi^0 \rightarrow \gamma \gamma$ amplitude.

The $K^0$-decay amplitude we study here has the  form
\begin{equation}
A(\overline{K^0} \rightarrow \gamma \gamma) \, =
 \, \sqrt{2} \, {\cal{R}} \,
 \epsilon_{\mu \nu \rho \sigma } \epsilon_1 ^{\mu} \epsilon_2^{\nu}
k_1^{\rho} k_2^{\sigma} \, ,
\end{equation}
 where the quantity $ {\cal{R}}$ can be split in two parts
as in (3).
The results which we presented in ref.\cite{ep93} imply that even in the
chiral limit $m_s \rightarrow 0$ there is a non-zero
contribution to ${\cal{R}}$ from
${\cal{L}}(s \rightarrow d)_{\gamma}$ of order ${\cal{O}}(p^4)$,
\begin{equation}
{\cal{R}} \, =
 \, - 4 h_B \, e_D \, B \,
 \, .
\end{equation}
Here $e_D = e Q_D = -e/3$ is the electric charge of  the down ($d,s$) quarks.
 The  quantity $h_B$ is the hadronic matrix element of the quark operator
within the parenthesis in (1) calculated
within an  effective low-energy  QCD described in \cite{leqcd,PdeR91}.
 This effective QCD has an extra term (4)
 which contains
meson-quark couplings proportional
to the constituent quark mass $M \sim$ 300 MeV and to
inverse powers of $f_\pi$. The chiral symmetry-breaking scale
 $\Lambda_{\chi} \sim 0.7$ to 1 GeV is the
natural UV cut-off.
 Potentially UV-divergent integrals $\sim \Lambda_{\chi}^2$
and $log(\Lambda_{\chi})$ are absorbed in the quark condensate
and the physical $f_\pi$, respectively. The result of the
calculation  is \cite{ep93}
\begin{equation}
 h_B \, = \, - \frac{N_c M^2 \Delta_{LD}}{4 \pi^2 f_{\pi}} \; .
\end{equation}
 The quantity $\Delta_{LD}$ is a dimensionless function of
the ratio  $M^2/\Lambda_{\chi}^2$.
 Using dimensional
regularization, corresponding to the formal limit
 $\Lambda_{\chi} \rightarrow \infty$, we find
 $\Delta_{LD} = -2$.
 The quantity $h_B$ in (6) is the sum of
two terms \cite{ep93}:
The one with the hadronic matrix element $\sim f_K$ obtained
 for $i D_\sigma \rightarrow e_D A_\sigma$
in (1) corresponding to irreducible electroweak diagrams  for $s \rightarrow
d \gamma \gamma$; the other term, having a hadronic matrix element
$\sim f_{LD} = f_K - h_B$, is obtained when $D_\sigma
\rightarrow \partial_\sigma$ in (1), and corresponds to an
irredicible diagram for
 $s \rightarrow d \gamma$ with a photon on an external
($s-$ or $d-$) quark,  i.e. to a reducible diagram for
 $s \rightarrow d \gamma \gamma$.
It should be noted that the total contribution
 $f_K - f_{LD} = h_B \rightarrow~0$ in the limit $f_\pi
\rightarrow \infty$ when the quark-meson interactions are switched
off, corresponding to the free-quark case.
There are also other operators, e.g. the non-diagonal $s \rightarrow d$
self-energy, but these turn out to have small coefficients and
do not contribute significantly to $K \rightarrow 2\gamma$.
\vspace{0.3cm}

In the following we concentrate on the CP-conserving case.
In order to explain $K_L \rightarrow \gamma \gamma$
by ${\cal{L}}(s\rightarrow d)_{\gamma}$ alone, a numerical value
$\hat{B}_u \simeq 1.5$
 is needed and pure electroweak diagrams such as in
\cite{sdgg} cannot account for this in the CP-conserving case.
However, some QCD-induced contributions exist.
A typical contribution, which has already been known
for some time  \cite{SVZg}, corresponds to
\begin{equation}
\hat{B}_u \, = \, 2 Q_U \frac{\alpha_s}{\pi} \ln
\frac{m_c^2}{\mu^2} \; ,
\end{equation}
where $Q_U = 2/3$ is the charge of up  (u,c) quarks, and $\mu$ is the
renormalization scale, typically of order 1 GeV. Taking this
expression at face value gives
 $\hat{B}_u \simeq 0.3$  for $\mu = \Lambda_\chi$
and $\hat{B}_u \simeq 1.2$
for $\mu = M$. However, the physical result corresponding to the
leading logarithmic perturbative result (8)
is difficult to estimate reliably because
the charm scale is rather close to the
confinement and the chiral-symmetry breaking scales (so that
next-to-leading-log terms might be important).
Taking into account that $\mu$ corresponds to the $s,d$
quark momenta in (1), subsequently being the loop momenta in the
effective low-energy QCD \cite{leqcd}, we find that
 $\hat{B}_u \simeq 0.6$ to 1
is a reasonable range for the value based on (8).
In previous papers \cite{ep88,ep89} we studied "penguin-induced"
two-loop contributions to $s \rightarrow d \gamma, d \gamma\gamma$
and $b \rightarrow s \gamma$,
where we focused on the $(c,t)$ quarks in the loop. There will
also be corresponding CP-conserving terms $\sim \lambda_u$ for
 $s \rightarrow d \gamma, d \gamma\gamma$
with $(u,c)$ quarks in the loops. These will, however, give too
small a contribution to $\hat{B}_u$.
There will also be contributions
of the non-diagonal self-energy type \cite{ep93} which are
non-zero, but small.

Taking into account the uncertainty
of perturbative calculations on the 1 GeV scale,
QCD-induced contributions will probably reproduce roughly
half of the $K_L \rightarrow \gamma \gamma$ decay amplitude.
These SD contributions require additional next-to-leading
corrections to (8) in order to reach a more precise conlusion.
In addition, there might be  a significant long distance contribution
from improper cancellation of pole diagrams.

\vspace{0.5cm}

\begin{Large}

3.  Anomalous nature of the
$K_L \rightarrow \gamma \gamma$ amplitude

\end{Large}

\vspace{0.2cm}

In order to establish the anomalous nature of the
$K_L \rightarrow \gamma \gamma$ process, we perform a comparative
study  of  $K_L \rightarrow \gamma \gamma$ and
$\pi^0 \rightarrow \gamma \gamma$ processes. Both of them
are governed by an effective interaction of the form
\begin{equation}
{\cal L}_X \, = \,
 { \alpha  C_X} \epsilon_{\mu\nu\rho\sigma}
F^{\mu\nu}F^{\rho\sigma} \Phi_X\; ,
\end{equation}
for $X= \pi^0$ or $X=K_2$.
The  $\pi^0 \rightarrow   \gamma \gamma$ rate is reproduced by
\begin{center}
$C_{\pi^0 } \, = \, {N_c \over 24 \pi f_\pi} \,= \, 4.3 \times 10^{-4}
MeV^{-1} \, .$
\end{center}
 The rate for $K_2 \rightarrow   \gamma \gamma$  similarly
determines the phenomenological coupling
\begin{center}
$|C_{K_2}| \, = \,
 5.9 \times 10^{-11} MeV^{-1}\, .$
\end{center}

Notably, the adopted low-energy QCD accounts for the full $C_{\pi^0}$
amplitude, whereas the calculation in the previous section shows that
it accounts for roughly a half of the $|C_{K_2}|$. This
refers to the "unrotated" ($U$) version of low-energy QCD \cite{leqcd}.

 However, the term ${\cal{L}}_{\chi}$ in (4) can be transformed into a pure
mass term $- M \bar{{\cal{Q}}} {\cal{Q}}$
for rotated "constituent quark" fields ${\cal{Q}}_{L,R}$ :
\begin{equation}
q_L \rightarrow  {\cal{Q}}_L =  \xi  q_L \; ; \;  \; \;
q_R \rightarrow  {\cal{Q}}_R =  \xi^\dagger q_R \;  ;
\; \; \xi \, \cdot \, \xi  = U \; .
\end{equation}
Then the meson-quark couplings in this "rotated" $(R)$ picture  are
transformed into the kinetic
(Dirac) part of the "constituent quark" lagrangian.
These interactions can be described in terms of vector and axial vector
fields coupled to constituent quark fields
${\cal{Q}} = {\cal{Q}}_R + {\cal{Q}}_L$:
\begin{eqnarray}
{\cal{L}}_{int} \, = \, \bar{{\cal{Q}}} [ \gamma^{\mu} {\cal{V}}_{\mu} \,  + \,
\gamma ^{\mu} \gamma _5   {\cal{A}}_{\mu} \, ] {\cal{Q}} \; \; ;  \nonumber \\
{\cal{V}}_{\mu} \, = \, (R_{\mu} + L_{\mu})/2  \; \; ;
{\cal{A}}_{\mu} \, = \, (R_{\mu} - L_{\mu})/2  \; \; ;  \nonumber \\
L_{\mu} \, = \, \xi \, (i \partial_{\mu} \, \xi^\dagger) +
 \xi \, l_\mu \, \xi^\dagger \; ; \;
R_{\mu} \, = \, \xi^\dagger (i \partial_{\mu} \xi) + \xi^\dagger
r_\mu \, \xi \; .
\end{eqnarray}
Here $l_\mu$ and $r_\mu$ are the external fields containing the photon
(eventually also the W field).
A calculation of the $\pi^0 \rightarrow \gamma \gamma$ amplitude
in the rotated ($R$) picture gives a zero result
(for infinite cut-off, i.e. by dimensional regularization).
We have found by explicit calculation that this is also the case
for the $\overline{K^0} \rightarrow \gamma \gamma$ amplitude!

 For the
$\pi^0 \rightarrow \gamma \gamma$  decay,
the interpretation is unambiguous. In the rotated basis,
where pions have proper derivative Goldstone-couplings,
the compensating WZW term ensures the anomaly matching.
The unrotated-quark-triangle evaluation finds a counterpart in the
anomalous WZW part of the chiral lagrangian.
The explicit diagrammatic evaluation complies with the more general
functional derivation of the Wess-Zumino-Witten
(WZW) term \cite{WZ}, which is contained in a Jacobian of the quark
field rotation in eq. (10).

The results
(non-zero in the $U$-basis, zero in the $R$-basis)
in evaluating the $\overline{K^0} \rightarrow  \gamma \gamma$
amplitude from  quark-level diagrams within low-energy QCD,
motivates us to attribute a similar anomalous nature to this process!
By employing the anomaly-matching principle, we argue for the
existence of the related bosonic lagrangian term corresponding to the
WZW term.

In this connection we should stress an important point, namely  that
the wanted bosonic lagrangian is based on the underlying
$s \rightarrow d \gamma \gamma$ transition that is not of the
current-current form explored in the literature  \cite{enp,bep,p92}.
In the chiral-invariant version,
the quark operator (1) can be rewritten in a form where its
$(8_L , 1_R)$ structure is manifest.
For convenience, we also perform the Dirac algebra in (1) in
order to bring it to the form containing $\sigma_{\mu \nu}$
instead of the Levy-Civita tensor
(which obscures the dimensional regularization):
\begin{equation}
{\cal{L}}(s\rightarrow d)_{\gamma} \, = \,  (-3) B \,
 \overline{q_L} \; \lambda_+ [ i \stackrel{\leftarrow}{D} \cdot  \gamma
 \; \sigma^{\mu \nu} \; F_{\mu \nu}^L  \; + \;
\sigma^{\mu \nu} F_{\mu \nu}^L \;
 i \gamma \cdot \stackrel{\rightarrow}{D}  ] \;  q_L  \; .
\end{equation}
In this expression, $D_{\sigma} q_L  = (\partial_{\sigma} - i l_{\sigma}) q_L
$,
$\bar{\psi} \stackrel{\leftarrow}{\partial_{\sigma}}
 \, = \, - \partial_\sigma (\bar{\psi})$, the field-strength tensor
$F^L_{\mu \nu} \, = \, \partial_\mu l_\nu - \partial_\nu l_\mu
- i\,  [l_\mu \, , \, l_\nu]$,
 and the factor  $(-3)$  compensates  the charge $Q$  contained
in the field $l_{\mu}$.  For the purely electromagnetic
gauging, this field is
\begin{center}
$l_\mu= r_\mu \equiv e A_\mu Q;
 \quad Q = \frac{1}{3} diag(2,-1,-1).$
\end{center}
The Gell-Mann matrices
$\lambda_{\pm} \, = \, (\lambda_6 \pm i \, \lambda_7)/2$ project
$\Delta S \, = \, \pm 1$ transitions  out of the quark fields
 $\bar{q}_L \, = \,
(\bar{u}_L, \, \bar{d}_L, \, \bar{s}_L )$.
 Note that in (12) the
covariant derivative contains only the left-handed field $l_\sigma$,
and that the operator transforms as $(8_L, 1_R)$ under
the local chiral $SU(3)_L\times SU(3)_R$ symmetry. Obviously, because of the
anomalous nature of $K^0 \rightarrow  2\gamma$, we have established,
the local chiral invariance should be lost, and we expect that our
amplitude is in some way related to the WZW term.

In ref. \cite{ep93} it is shown that the term
${\cal{L}}(s\rightarrow d)_{\gamma}$ can be transformed away into
the kinetic QCD term. However, compensating interactions are then
appearing in the "mass term" (4), and can be described by
a new lagrangian, which has an appealing form after performing the
rotation (10) (which simultaneously gives (11)) :
\begin{equation}
\Delta {\cal{L}}_\chi^{\Delta S=1} \, = \,  (-3) B \, M \,
 \overline{{\cal{Q}}} \;
\sigma_{\mu \nu} \; T_{L (+)}^{\mu\nu}  \; {\cal{Q}}  \; ,
\end{equation}
where
\begin{equation}
 T_{L (+)}^{\mu\nu} \, = \xi \, \lambda_+ \, F_{\mu \nu}^L \xi^\dagger \;
\end{equation}
encapsulates the flavour change in the photon-emission vertex.
Note that also from $\Delta {\cal{L}}_\chi^{\Delta S=1}$ we have
found by explicit calculation, using dimensional regularization, that
$\Delta_{LD} = -2$ and $\Delta_{LD} =0$ in the $U$- and $R$- pictures
respectively (see eq.(7)), thus confirming the anomalous
charachter of $K^0 \rightarrow  2\gamma$. In the next section we will
show how the expressions (11) and (13) enable us to
find the general structure
of the hadronic   ${\cal{O}}(p^4)$ term responsible for
the  $K^0 \rightarrow  2\gamma$ process.

\vspace{0.5cm}

 \begin{Large}
4. Bosonization
\end{Large}

\vspace{0.2cm}

Our final task refers to
" bosonizing" the operator contributing to
 $K^0 \rightarrow  2\gamma$, induced by (1) or (12)  or by
"the intermediate" expression, which still includes the
Levy-Civita tensor:
\begin{equation}
{\cal{L}}(s\rightarrow d)_{\gamma} \, = \,  (-3) B \,
\epsilon^{\mu \nu \sigma \rho}
 \bar{q_L} \; \lambda_+ (i \stackrel{\leftarrow}{D_{\sigma}}
\; F_{\mu \nu}^L \, + \, F_{\mu \nu}^L
 \;i \stackrel{\rightarrow}{D_{\sigma}} ) \gamma_{\rho} q_L   \; .
\end{equation}
A direct bosonization of this  operator is difficult because of
the covariant derivative sitting between the quark fields.
However, we observe that by keeping only the $l_\mu$ part of the
covariant derivative
(corresponding to purely electromagnetic gauging),
we obtain the promising structure
\begin{equation}
\epsilon^{\mu \nu \sigma \rho} \;
( {\em quark \; current} )_{\rho}  \times l_\sigma
\times F^L_{\mu \nu}.
\end{equation}
The bosonized version of the quark current
is  $\Sigma^{L}_{\rho} =  U^{\dagger} \partial_\rho U$, the object
which can be recognized as
a building block of the WZW terms responsible for radiative decays
(-see eq.(19) below).

An elegant way to bosonize quark operators, the heat kernel method,
 is  demonstrated in \cite{PdeR91}. In the present paper we are,
however, sticking to quark loop diagrams which give less information
on the bosonized counterpart of ${\cal{L}}(s\rightarrow d)_{\gamma}$
in eqs.(1), (12) and (15).
Indeed, in the $U$- picture loop calculation,
the momenta which correspond to derivatives on fields have to be
 pulled out of the quark propagators of the loop diagram,
 and we loose  structures like (16). However,
this is not so in the $R$- picture, where the vector and
axial vector fields in eq.(11) couple in two of the vertices of
the triangle loop diagram for $\overline{K^0} \rightarrow 2 \gamma$, and
the interaction given by (13) acts in the third vertex. Then, for
${\cal{O}}(p^4)$ terms, no momenta have to be pulled out of the
quark propagators, and
we can deduce the typical anomalous contribution:
\begin{equation}
{\cal L}_{An}^{\Delta S=1} \sim
\frac{M^2 N_c}{4\pi^2 f_{\pi}} \; B \;
 \epsilon^{\mu\nu\alpha\beta}~
 Tr \; [{T_L^{(+)}}_{\mu\nu} {\cal{V}}_{\alpha} {\cal{A}}_{\beta}]\; .
\end{equation}
There will also be other terms with the structure
${\cal{V}} {\cal{A}}$ replaced by
${\cal{A}} {\cal{V}}$, ${\cal{V}} {\cal{V}}$, or
${\cal{A}} {\cal{A}}$,
which will appear in the total lagrangian with different coefficients.
These coefficients have to be determined in the $U$-picture, by
considering quark loop diagrams for processes to which (17)
contributes.
It is known \cite{leqcd} that
${\cal{A}}_\mu$ is
locally invariant, whereas the vector field ${\cal{V}}_\mu$ is not.
Therefore, the terms involving the vector field manifestly
break the local chiral invariance in a transparent way.
We will discuss  expression (17) in full detail elsewhere.
Here we explicate this
formula for the $K^0 \rightarrow  2\gamma$ process by comparing it
to a more familiar WZW term:   Operators like (17) might be obtained
 by appropriate $\lambda_+$ insertions in the WZW action.

Some comments on the ordinary Wess-Zumino-Witten
(WZW) term \cite{WZ} are in order. Apart from the hadronic term
(relevant to $K^0 \overline{K^0} \rightarrow 3\pi$ )  there is a part
relevant to the Goldstone-to-two-photon
radiative decays under consideration:
\begin{equation}
{\cal L}_{WZW}= -{i N_c \over 48\pi^2}\epsilon^{\mu\nu\alpha\beta}~
 Tr \; V_{\mu\nu\alpha\beta}\; ,
\end{equation}
where
\begin{equation}
V_{\mu\nu\alpha\beta} \, = \,
\Sigma^{L}_{\mu} U^\dagger\partial_\nu r_\alpha U l_\beta
+ \Sigma^{L}_{\mu} l_\nu \partial_\alpha  l_\beta
+ \Sigma^{L}_{\mu} \partial_\nu l_\alpha  l_\beta \, + ....
- ( L \leftrightarrow R ) \; .
\end{equation}
The explicit  $ L \leftrightarrow R$ symmetry interchanging $U$
and $ U^{\dagger}$, and $\Sigma^{L}_{\mu} =  U^{\dagger} \partial_\mu U$ and
$\Sigma^{R}_{\mu}  = U  \partial_\mu U^{\dagger} $,
\begin{center}
$U \leftrightarrow  U^\dagger \;  ;
\Sigma^{L}_{\mu} \leftrightarrow \Sigma^{R}_{\mu} \; ,$
\end{center}
leads to doubling of the "odd $\Pi$" terms and cancelling of the "even $\Pi$"
terms.

At  order ${\cal O}(p^4)$ this gives the famous term responsible for
 $\pi^0 \rightarrow \gamma \gamma$, which does not involve any unknown
coefficients:
\begin{eqnarray}
{\cal L}^{(4)}_{WZW} \, = \,
 {N_c \alpha \over 24 \pi f_\pi} \epsilon_{\mu\nu\rho\sigma}
F^{\mu\nu}F^{\rho\sigma} \biggl(\pi^0 + {\eta \over \sqrt{3}}
\biggr) \nonumber \\
\, - \, { i  N_c e \over 12 \pi^2f_\pi^3} \epsilon_{\mu\nu\rho\sigma} A^\mu
\partial^\nu \pi^+ \partial^\rho \pi^- \partial^\sigma\biggl(\pi^0 +
{\eta \over \sqrt{3}}\biggr) \;  + .... .
\end{eqnarray}
Note that the first term in (20 ) acquires the form written above after
applying partial integration to the derivative of the meson field!
Before this partial integration
the first term had the form
\begin{center}
$\epsilon_{\mu\nu\rho\sigma}
F^{\rho\sigma}A^{\nu} \partial^{\mu}(\pi^0 + \eta/\sqrt{3})$. \\
\end{center}

The two terms in (20) participate in the
{\em reducible} anomalous (long-distance  pole) contributions
to $K_L \rightarrow 2\gamma$ and $K_L \rightarrow \pi \pi \gamma$,
respectively.  The other vertex in such pole graphs is given
by the weak $K \rightarrow \pi$ transition, displayed below in eq. (23).
The fact that the $\pi^0$ and $\eta$ pole contributions cancel  owing to the
Gell-Mann-Okubo mass relation $m^2_\eta={1\over 3}(4m^2_K-m^2_\pi)$
arouses  interest in the possible existence of
 important non-pole contributions.

Thus, the question we raise here concerns the possibility of
  "flavour extension" of
 expression (19) to the $\Delta S = 1$ case, i.e.  the possibility of the
corresponding expression with $\partial^\sigma \pi^0$ replaced by
$\partial^\sigma \overline{K^0}$.
We observe that the simple substitution
\begin{center}
$ 1 \rightarrow {\cal{G}}_{\rm WZW} \, \lambda_+$
\end{center}
in front of $V_{\mu\nu\alpha\beta}$ in (18) provides us with such
terms.
The lagrangian obtained in this way,
\begin{equation}
{\cal{L}}^{\Delta S=1}_{\rm WZW} \, = \,  -{i N_c \over 48\pi^2}
{\cal{G}}_{\rm WZW} \,
 \epsilon^{\mu\nu\alpha\beta}
\, Tr(\lambda_+ V_{\mu\nu\alpha\beta}) \; ,
\end{equation}
and the bosonized form ${\cal{L}}^{\Delta S=1}_{An}$ in (17)
both contain the representative terms for
 $\overline{K^0} \rightarrow 2\gamma$.
In general, this prescription is not unique because the $\lambda_+$
can be inserted in several ways inside the tensor
 $V_{\mu \nu \alpha \beta}$ in (19). Thus (21) represents a rather schematic
notation: There are different  ${\cal{G}}_{\rm WZW}$ couplings for different
insertions of $\lambda_+$.  A discussion on the different terms
of (21) and its relation to (17) will be given elsewhere.

 The prescription
used to obtain (21) repeats "the Cronin's substitution"
 leading from the ${\cal{O}}(p^2)$
strong/electromagnetic term
\begin{equation}
{\cal{L}}^{(2)}_{strong/em} \, = \, \frac{f_{\pi}^2}{4} \,
 Tr(D^{\mu} U D_{\mu} U^{\dagger}) \; ,
\end{equation}
to the corresponding non-leptonic $\Delta S = 1$ term \cite{cro67}
\begin{equation}
{\cal{L}}^{(2)}_{\Delta S = 1} \, = \, g_8 \,
 Tr(\lambda_+ D^{\mu} U D_{\mu} U^{\dagger}) \; .
\end{equation}
In close correspondence with the flavour-diagonal terms in (14),
 our new anomalous term(s) ${\cal{L}}^{\Delta S=1}_{\rm WZW}$
accounts for the transitions
$\overline{K^0} \rightarrow  \gamma \gamma,
\pi \pi \gamma, \pi \pi \gamma \gamma$, but not for
$\overline{K^0} \rightarrow  \pi^0 \gamma \gamma$ :
\begin{eqnarray}
{\cal{L}}^{\Delta S=1}_{\rm WZW}{(4)}\, = \,
{\cal{G}}_{\rm WZW} {N_c \alpha \over 24 \pi f_\pi} \frac{2 \sqrt{2}}{3}
 \epsilon_{\mu\nu\rho\sigma}
A^{\nu}F^{\rho\sigma} \partial^{\mu}\overline{K^0}
 \nonumber \\
-  {\cal{G}}'_{\rm WZW} { N_c e \over 12 \pi^2f_\pi^3} \frac{\sqrt{2}}{3}
\epsilon_{\mu\nu\rho\sigma}
A^{\sigma} \partial^{\mu}\overline{K^0}
[\partial^{\nu}\pi^- \partial^{\rho}\pi^+]
\nonumber \\
+  {\cal{G}}''_{\rm WZW}
(\overline{K^0} \rightarrow  \pi \pi \gamma \gamma)_{term} \; .
\end{eqnarray}
Note that the $\lambda_+$ -insertion responsible for the first term
in (24) is unique.  For the other terms, however,  it is not unique ;
the $\lambda_+$
can be inserted at several places within expression (19).
Thus we focus on  determining the coupling ${\cal{G}}_{\rm WZW}$
 from the uniquely parametrized $K^0$-decay
into two photons - provided that this term in (24) explains the total
${K_L} \simeq K_2 \rightarrow
\gamma \gamma$  rate.

Neglecting CP-violation, we can determine
${\cal{G}}_{\rm WZW}$ from the  measured
$K_2  \gamma \gamma$ coupling $C_{K_2}$ in
eq.(9):
\begin{equation}
|C_{K_2}| \, = \, {2 \over  3} \, |{\cal{G}}_{\rm WZW}| \, C_{\pi^0 } \;
\; \; \;  ;   \; \; \; \; \; \;
 |{\cal{G}}_{\rm WZW}| \, \simeq  \,  2 \times 10^{-7} \; .
\end{equation}
A sizable part of this coupling is reproduced by a quark-loop evaluation
of Sect.2. It reflects a share of the anomaly contribution in the
effective coupling ${\cal{G}}_{\rm WZW}$ ,
which comprises all possible short- and long-distance contributions,
and as such has to be determined outside $\chi$PT.

The second term in eq.(24) contributes to the decay
$K_L \rightarrow \pi^+  \pi^- \gamma$. Calculation within the
low-energy QCD\cite{leqcd} shows that
${\cal{G}}'_{\rm WZW}$ is suppressed by $M^2/\Lambda_{\chi}^2$ with
respect to ${\cal{G}}_{\rm WZW}$.

\vspace{0.5cm}

\begin{Large}

5.Discussion and Conclusions

\end{Large}

\vspace{0.2cm}

We have presented a new non-negligible contribution to $K_L \rightarrow
\gamma \gamma$ obtained within the effective low-energy QCD from
a short-distance operator.
This contribution can account for about half of the
experimental value, within the leading logarithmic approximation.
Within the standard nomenclature \cite{enp}
at the hadronic level, this process should belong to the class of
{\em reducible anomalous} pole diagrams $K_L\rightarrow \pi^0,
\eta \rightarrow \gamma \gamma$, which have so far been
considered to be the only contribution of ${\cal O}(p^4)$.
As already stated, this amplitude is subjected to pole cancellation,
so that only some   ${\cal O}(p^6)$ contribution should remain.
The real representative of  ${\cal O}(p^4)$
{\em reducible anomalous} neutral kaon decays remains
\begin{center}
 $K_L \rightarrow \pi^+ \pi^-
\gamma \gamma \; .$
\end{center}
How does  our  ${\cal O}(p^4)$ contribution to $K_L\rightarrow
 \gamma \gamma$ compare with the existing list of
{\em direct} anomalous processes \cite{enp}?
The origin of these direct anomalous terms  can be traced back by
following the authors of refs. \cite{enp,bep,p92}.
They  started from a four-quark
effective hamiltonian essentially involving products of two weak
currents (in the factorizable limit), using the functional derivative
of the action $S=\int d^4 x {\cal{L}}$ as an identification of the
quark current:
\begin{equation}
\bar{q} \gamma^{\mu} L q \sim  \frac{\delta S}{\delta l_{\mu}}  \; .
\end{equation}
 To obtain the anomalous terms,
the authors of refs. \cite{enp,bep,p92} wrote down an
expression  of the form
\begin{equation}
 {\cal{L}}_{\rm An} \, \sim G_F \,
\biggl( \frac{\delta S^{(2)}}{\delta l_{\mu}} \biggr)
\; \biggl( \frac{\delta S_{WZW}}{\delta l^{\mu}} \biggr) \; ,
\end{equation}
where $S^{(2)}$ is the normal action ${\cal{O}}(p^2)$.
The result
did not contribute to
$\overline{K^0} \rightarrow \gamma \gamma$ because
a chiral-invariant expression typically contains
(in the elecromagnetic case $l_\mu = r_\mu =  e Q A_\mu$)
the covariant derivative of $U$,
$D_{\mu} U=\partial_\mu U-ieA_\mu[Q,~U]$, or U in the combination
$U Q U^\dagger = Q + [\Pi,Q] + ... \;  .$ It is easily seen that
 the commutator
$[\Pi,Q]$ does not contain neutral kaons ($Q$ acts as the unit matrix
in the $s, d$ sector).
 To find a contribution giving the $\overline{K^0} \rightarrow \gamma
\gamma$ amplitude in this way, one has to go further to
  ${\cal{O}}(p^6)$ terms.
This should be no surprise when looking at the underlying quark processes.
In the CP-conserving sector, where the CKM-favourable $s \rightarrow d
\gamma \gamma$ has $u$  and $c$ quarks running in the loop, the SD
contributions giving rise to ${\cal{O}}(p^4)$ terms are effectively
cancelled by the GIM mechanism. Thus the neutral kaon decay
representative of this {\em direct} anomalous class , which can be read off
in  refs. \cite{enp,bep}, is
\begin{center}
 $K_L \rightarrow \pi^+ \pi^-
\gamma  \; .$
\end{center}
Our assertion is that our new
WZW-extended $\Delta S = 1$ term
adds new "odd $\Pi$" processes to the existing list of
{\em direct} anomalous processes \cite{enp},
namely the decay
\begin{center}
 $K_L \rightarrow
\gamma  \gamma  \; .$
\end{center}
We have arrived at
additional  a priori legitimate operators to the effective
theory of the $\chi$PT type
by integrating out the quark loops in an effective low-energy QCD.
 A guideline in introducing them are bosonized
forms of the operators in the underlying quark theory. The new terms we
have suggested rely on
two-quark operators,  which have not been considered as yet.
Our contributions, termed  "off-shell contributions" in ref.  \cite{ep93},
were obtained  in departing from the free-quark picture,
in compliance with low-energy QCD.
By scrutinizing the appearance of the anomaly in the U- and R-versions
of  low-energy QCD,  we demonstrate both the viability of
the anomaly-matching principle  and the kinship of the
 $K_L \rightarrow \gamma \gamma$ and
$\pi^0 \rightarrow \gamma \gamma$ decays .
The weak flavour transition, which distinguishes these two processes,
is merely a decoration on the top of the dynamics underlying the anomaly.
We hope that further investigation in this direction might enlighten
the role of the chiral anomaly in flavour-changing radiative transitions
and/or decode  some secrets of bosonization in the Goldstone boson sector.

\vspace{0.7cm}

{\bf Acknowledgement}

\vspace{0.1cm}

One of us (J.O.E.) wants to thank H. Bijnens,  W. Bernreuther, R. Decker,
G. Ecker, H. Neufeld and H. Pilkuhn for useful comments.
I.P. acknowledges
the hospitality of the Fysisk Institutt (Universitetet i Oslo),
the partial support
of the contracts CI1*-CT91-0893 (HSMU) and JF-899-31/NSF,
and helpful discussions with
H. Bijnens, G. Ecker and D. Klabu\v{c}ar.

\end{document}